\documentstyle[11pt,epsfig,amsfonts]{article}

\def\beq{\begin{equation}}
\def\eeq{\end{equation}}
\def\beqa{\begin{eqnarray}}
\def\eeqa{\end{eqnarray}}

\def\eqn#1{Eq.~(\ref{#1})}
\def\eqns#1#2{Eqs.~(\ref{#1}) and~(\ref{#2})}
\def\eqnss#1#2{Eqs.~(\ref{#1})--(\ref{#2})}
\def\fig#1{Fig.~{\ref{#1}}}
\def\sec#1{Section~{\ref{#1}}}

\def\tab#1{Table~\ref{#1}}

\newcommand\sss{\scriptscriptstyle}
\def\abs#1{\left| #1\right|}
\def\dy{\Delta y}
\def\de{{\cal D}}

\def\cF{{\cal F}}
\def\cFt{\tilde{\cal F}}
\def\cG{{\cal G}}
\def\cGt{\tilde{\cal G}}
\def\Cuts{{\cal C}}
\def\hs{\hat s}
\def\mn{{\sss {\rm MN}}}
\def    \kta              {\mbox{$k_{a\perp}$}}
\def    \ktb              {\mbox{$k_{b\perp}$}}
\def    \ktat             {\mbox{$k_{a\perp}^2$}}
\def    \ktbt             {\mbox{$k_{b\perp}^2$}}
\def    \vkta             {\mbox{$\vec{k}_{a\perp}$}}
\def    \vktb             {\mbox{$\vec{k}_{b\perp}$}}
\def    \qta              {\mbox{$q_{a\perp}$}}
\def    \qtb              {\mbox{$q_{b\perp}$}}
\def    \qtat             {\mbox{$q_{a\perp}^2$}}
\def    \qtbt             {\mbox{$q_{b\perp}^2$}}
\def    \Ecut             {\mbox{$E_\perp$}}
\def    \Ecutt            {\mbox{$E_\perp^2$}}
\def    \muf              {\mbox{$\mu_{\sss F}$}}
\def    \muft             {\mbox{$\mu_{\sss F}^2$}}

  \newcommand{\ccaption}[2]{
    \begin{center}
    \parbox{0.85\textwidth}{
      \caption[#1]{\small{\it{#2}}}
      }
    \end{center}
    }

\makeatletter
\def\@eqnnum{\hbox{\reset@font\rm(\theequation)}}
\let\make@eqnnum=\@eqnnum %
\def\eqnum#1{\dec@eqnnum \global\def\make@eqnnum{\reset@font\rm(#1)}%
\def\@currentlabel{#1}%
}
\def\inc@eqnnum{\addtocounter{equation}{1}}
\def\dec@eqnnum{\addtocounter{equation}{-1}}
\@definecounter{equation}%
\@addtoreset{equation}{section} %
\def\theequation@prefix{{\thesection}.} %
\def\theequation{\theequation@prefix\arabic{equation}}%
\makeatother

\begin{document}

\begin{titlepage}

\hspace*{\fill}\parbox[t]{4cm}{
IPPP/00/04 \\
DTP/00/64\\
MSUHEP-01024\\
DFTT 44/2000\\
\today}

\vspace{1.cm}

\begin{center}
{\Large\bf Mueller-Navelet Jets at Hadron Colliders}\\
\vspace{1.cm}

{J.~R.~Andersen$^1$, V. Del Duca$^2$, S. Frixione$^3$, C.R. Schmidt$^4$
and W.J. Stirling$^1$}\\
\vspace{.2cm}
{$^1$\sl Institute for Particle Physics Phenomenology\\ 
University of Durham\\
Durham, DH1 3LE, U.K.}\\

\vspace{.2cm}
{$^2$\sl I.N.F.N., Sezione di Torino\\
via P. Giuria, 1 - 10125 Torino, Italy}\\

\vspace{.2cm}
{$^3$\sl I.N.F.N., Sezione di Genova\\
via Dodecaneso, 33 - 16146 Genova, Italy}\\

\vspace{.2cm}
{$^4$\sl Department of Physics and Astronomy\\
Michigan State University\\
East Lansing, MI 48824, USA}\\

\vspace{.5cm}

\begin{abstract}
We critically examine the definition of dijet cross sections at
large rapidity intervals in hadron--hadron collisions, 
taking proper account of the various
cuts applied in a realistic experimental setup. We argue that the
dependence of the cross section on the precise definition of the
parton momentum fractions and the
presence of an upper bound on the momentum transfer cannot be
neglected, and we provide the relevant modifications to the
analytical formulae by Mueller and Navelet. We also point out
that the choice of equal transverse momentum cuts on the tagging
jets can spoil the possibility of a clean extraction of signals
of BFKL physics.
\end{abstract}

\end{center}
 \vfil

\end{titlepage}

\section{Introduction}
\label{sec:intro}

Long ago Mueller and Navelet suggested~\cite{Mueller:1987ey} to look for
evidence of  Balitsky-Fadin-Kuraev-Lipatov (BFKL) 
evolution~\cite{Kuraev:1976ge,Kuraev:1977fs,Balitsky:1978ic}
by measuring dijet cross section at hadron colliders as a function of the 
hadronic centre-of-mass energy $\sqrt{S}$, at fixed momentum fractions 
$x_{a,b}$ of the incoming partons. This is equivalent to measuring the
rates as a function of the rapidity interval $\dy= |y_a-y_b|$ between 
the jets. In fact, at large enough rapidities, 
the rapidity interval is well approximated by the expression
$\dy\simeq \ln(\hs/|\hat t|)$, where $\hs=x_a x_b S$ and 
$|\hat t|\simeq \kta\ktb$, with $k_{a,b\perp}$ being the moduli of
the transverse momenta (i.e., the transverse energies) of the two jets. 
Thus, since the cross section tends to peak at the smallest available 
transverse energies, $\dy$ grows as $\ln S$ at fixed $x_{a,b}$.

It is clear that the measurement proposed by Mueller and Navelet is not
feasible at a collider run at a fixed energy; on the other hand, 
to look for the BFKL-driven rise of the parton cross section 
directly in the dijet production rate 
$d\sigma/\dy$ as a function of $\dy$
is difficult due to the steep fall-off of the parton 
densities~\cite{DelDuca:1994mn,Stirling:1994zs}. 
This led some of us to propose the study of less inclusive observables.
In particular, the angular correlation between the tagged jets, which at 
leading order (LO) are back-to-back, is smeared by gluon radiation 
and by hadronization. Part of the gluon radiation originates from
the mechanism responsible for BFKL effects, namely from gluon radiation 
in the rapidity interval between the jets. Accordingly, the transverse 
momentum imbalance~\cite{DelDuca:1994mn, DelDuca:1994xy} and the azimuthal
angle decorrelation~\cite{DelDuca:1994mn,Stirling:1994zs,DelDuca:1995fx,
DelDuca:1995ng,Orr:1997im} have been proposed as observables sensitive 
to BFKL effects. The azimuthal angle decorrelation has indeed been
studied by the D0 Collaboration at the Tevatron
Collider~\cite{Abachi:1996et}. 
As expected, a NLO partonic Monte Carlo generator,
JETRAD~\cite{Giele:1993dj,Giele:1994gf}, predicts too little
decorrelation.  
However, the BFKL formalism predicts a
much stronger decorrelation than that observed in the data.  
In fact, the data are well described by the HERWIG
Monte Carlo generator~\cite{Marchesini:1988cf,Knowles:1988vs,
Marchesini:1992ch}, which dresses the basic $2\to 2$ parton scattering 
with parton showers and hadronization. This hints at a description of
the azimuthal angle decorrelation in terms of a standard Sudakov
resummation~\cite{Dokshitzer:1997uz}. It therefore appears that, in
the presence of Sudakov logarithms, it is quite difficult to cleanly
extract the presence of BFKL logarithms from this observable, 
not least because the latter are expected to be smaller than the former 
in the energy range
explored at present. It thus comes as no surprise that D0
Collaboration~\cite{Abachi:1996et} find no strong evidence of BFKL effects in
their data.

Recently, the D0 Collaboration~\cite{Abbott:2000ai} has revisited 
the original Mueller-Navelet proposal, and has measured the ratio
\beq
R = {\sigma(\sqrt{S_{_A}}) \over \sigma(\sqrt{S_{_B}})}
\label{ratio}
\eeq
of dijet cross sections obtained at two different centre-of-mass
energies, $\sqrt{S_{_A}} = 1800$ GeV and $\sqrt{S_{_B}} = 630$ GeV.
The dijet events have been selected by tagging the most forward/backward 
jets in the event, and the cross section is measured as a function
of the momentum transfer, defined as $Q^2=\kta\ktb$, and of the
quantities
\beq
x_1 = \frac{2\kta}{\sqrt{S}} e^{\bar y} \cosh (\dy/2)\,,\qquad
x_2 = \frac{2\ktb}{\sqrt{S}} e^{-\bar y} \cosh (\dy/2)\, ,
\label{eq:one}
\eeq
with $\bar y = (y_a+y_b)/2$, $\dy = y_a-y_b \ge 0$, and $y_a$ ($y_b$) 
are the rapidities of the most forward (backward) jet. The dimensionless
quantities $x_1$ and $x_2$ 
are reconstructed from the tagged jets using~\eqn{eq:one}, irrespective 
of the number of additional jets in the final state. In leading-order kinematics, 
for which only two (back-to-back) jets are present in the final state, 
we have $x_1=x_a$ and $x_2=x_b$, the momentum fractions
of the incoming partons. Higher-order corrections entail that 
these equalities no longer hold; however, $x_1=x_a$ and $x_2=x_b$ are still 
reasonable approximations (unless one goes too close to the borders of  
phase space).  This implies that when the ratio in Eq.~(\ref{ratio})
is computed at fixed $x_1$ and $x_2$, the contributions due to the parton
densities cancel to a large extent, thus giving the possibility of studying
BFKL effects without any contamination from long-distance phenomena.

\begin{table}
\begin{center}
\begin{tabular}{|l||c|c|} \hline
& $x_1$ range & $x_2$ range
\\ \hline\hline
bin 1
  &0.06-0.10 & 0.18-0.22
\\ \hline
bin 2
  &0.10-0.14 & 0.14-0.18
\\ \hline
bin 3
  &0.10-0.14 & 0.18-0.22
\\ \hline
bin 4
  &0.14-0.18 & 0.14-0.18
\\ \hline
bin 5
  &0.14-0.18 & 0.18-0.22
\\ \hline
bin 6
  &0.18-0.22 & 0.18-0.22
\\ \hline
\end{tabular} 
\end{center}                                                            
\ccaption{}{\label{tab:bins}
($x_1,x_2$) bins, with the upper bound of range in $x_1$ not larger 
than the upper bound of the range in $x_2$.
}
\end{table}                                                               
In the analysis performed by D0~\cite{Abbott:2000ai}, jets have been
selected by requiring $k_{a,b\perp}>20$~GeV, $|y_{a,b}|<3$, and $\dy>2$.
The cross section was measured for ten ($x_1,x_2$) bins, of which we
list in \tab{tab:bins} the six with the upper bound of the range in $x_1$ 
smaller than or equal to the upper bound of the range in $x_2$ (the others 
may be obtained by interchanging $x_1\leftrightarrow x_2$). Finally, a cut
on the momentum transfer, \mbox{$400<Q^2<1000$~GeV$^2$}, has been imposed.
With larger statistics, a binning in $Q^2$ would also be possible.
These cuts select dijet events at large rapidity intervals. To have
a crude estimate of the typical $\dy$ values involved, we observe
that, in a given $(x_1,x_2)$ bin, the data accumulate at the minimum 
$x_1$ and $x_2$ in order to maximise the parton luminosity, and at 
minimum $k_\perp$ in order to maximise the partonic cross section. We 
can then use the LO kinematics~(\ref{eq:one}) to obtain the effective 
rapidity interval. For instance, in bin 5 we find \mbox{$\dy_{_A}\simeq 5.3$}  
at \mbox{$\sqrt{S_{_A}}$=1800 GeV}, and \mbox{$\dy_{_B}\simeq 3.1$} at 
\mbox{$\sqrt{S_{_B}}=630$~GeV}. In addition, we see that in the
large--$\dy$ limit, $\dy_{_A} \to \dy_{_B} + \ln(S_{_A}/S_{_B})$.

The data collected by D0 are  compared to BFKL predictions
as given by Mueller and Navelet, and an effective `BFKL intercept' is then
extracted (see Eq.~(\ref{asympsol}) below). 
However, we argue in this paper that the different reconstruction of the $x$'s
used by D0 as compared to the original Mueller-Navelet analysis
(see \eqns{eq:one}{mnkin}), and some of the acceptance cuts imposed in the 
experimental analysis, like the introduction of an upper bound on the momentum 
transfer $Q^2$, actually spoil the correctness of this procedure, and
require modifications of the Mueller-Navelet formulae. These modifications 
are subleading from the standpoint of the BFKL theory, however they have an 
impact on the extraction of the BFKL intercept at subasymptotic energies. 
Furthermore, the fact that 
dijet events are selected by means of transverse momentum cuts which
are the {\it same} for the two tagged jets poses additional problems: large
logarithms of (non-BFKL) perturbative origin enter the cross section, 
and thus the ratio of Eq.~(\ref{ratio}) is 
affected by the same kind of problems as the azimuthal decorrelation.

In this study we address the quantitative importance of these issues
on the D0 analysis using a combination of analytic and numerical
techniques and several different theoretical approximations
 for dijet production.
The paper is organized as follows: in Section~\ref{sec:naiveBFKL},
we study the impact of a different definition of the $x$'s and of
the upper bound on $Q^2$ in the framework of
the naive BFKL equation, and compare the result with the standard
Mueller-Navelet analysis. In Section~\ref{sec:MCBFKL},
our BFKL predictions are refined by considering running-coupling
effects, and implementing energy-momentum conservation; this is done by 
using Monte Carlo techniques. In Section~\ref{sec:equalcut}
the problem of equal transverse momentum cuts is investigated;
in Subsection~\ref{sec:FO} we use a fixed-order perturbative
computation (accurate to NLO), and in Subsection~\ref{sec:BFKLcuts}
we return to the BFKL formalism. Finally, Section~\ref{sec:concl} reports 
on our conclusions. 

\section{Dijet production at fixed $x$'s in the {\sl naive} BFKL approach}
\label{sec:naiveBFKL}

In large-rapidity dijet production, the rapidity interval is related to
the kinematic invariants through $\dy\simeq\ln(\hat s/|\hat t|)$, with
$\hat s$ the squared parton centre-of-mass energy and $|\hat t|$ of the
order of the squared jet transverse energy. Thus, in computing the dijet
production rate we encounter large logarithms, which arise in a 
perturbative calculation at each order in the coupling constant $\alpha_{s}$.
The large logarithms can be resummed in leading-logarithmic (LL)
approximation through the BFKL equation.

In the high-energy limit, $\hat s\gg |\hat t|$, the BFKL theory assumes 
that any scattering
process is dominated by gluon exchange in the crossed channel.
This constitutes the leading (LL) term of the BFKL resummation.
The corresponding QCD amplitude
factorizes into a gauge-invariant effective amplitude formed by 
two scattering centres, the LO impact factors, connected by the gluon 
exchanged in the crossed channel. The BFKL equation then resums the universal
LL corrections, of 
${\cal O}(\alpha_s^n\ln^n(\hat s/|\hat t|))$,
to the gluon exchange in the crossed channel. These are obtained in the
limit of strong rapidity ordering of the emitted gluon radiation, i.e.
for $n+2$ gluons produced in the scattering,
\beq
y_a\gg y_1\gg y_2\gg \ldots \gg y_{n-1}\gg y_n\gg y_b\, .\label{mrk}
\eeq

In dijet production at large rapidity intervals,
the crossed-channel gluon dominance~\footnote{The
crossed-channel gluon dominance is also used as a diagnostic tool for 
discriminating between different dynamical models for parton scattering.
In the measurement of dijet angular distributions,
models which feature gluon exchange in the crossed channel, like QCD, 
predict a characteristic $\sin^{-4}(\theta^\star/2)$
dijet angular distribution~\cite{Ellis:1992qq,Abe:1992sj,Weerts:1994ui}, 
while models featuring contact-term interactions, which do not have gluon 
exchange in the crossed channel,
predict a flattening of the dijet angular distribution
at large $\hat s/|\hat t|$~\cite{Abe:1996mj,Abbott:1998nf}.} ensures that
the functional form of the QCD amplitudes for gluon-gluon, gluon-quark 
or quark-quark scattering at LO is the
same; they differ only by the colour strength in the parton-production
vertices.  We can then write the cross section in the following 
factorized form~\cite{Mueller:1987ey}
\beq
{d\sigma\over dx_ a^0 dx_b^0}\, =\, \int d\ktat d\ktbt 
f_{\rm eff}(x_a^0,\muft)\, f_{\rm eff}(x_b^0,\muft)\,
{d\hat\sigma_{gg}\over d\ktat d\ktbt}\, ,\label{mrfac}
\eeq
where $\muf$ is the factorization scale, and $x_a^0,\, x_b^0$ are the 
parton momentum fractions~(\ref{eq:one}) in the high-energy limit,  
\beq
x_a^0 = \frac{\kta}{\sqrt{S}} e^{y_a}\qquad \qquad
x_b^0 = \frac{\ktb}{\sqrt{S}} e^{-y_b}\, ,
\label{nkin0}
\eeq
and the effective parton densities are~\cite{Combridge:1984jn} 
\beq
f_{\rm eff}(x,\muft) = G(x,\muft) + {4\over 9}\sum_f
\left[Q_f(x,\muft) + \bar Q_f(x,\muft)\right], \label{effec}
\eeq
where the sum runs over the quark flavours. In the high-energy limit, 
the azimuthally-averaged gluon-gluon scattering cross section 
becomes~\cite{Mueller:1987ey}
\beq
{d\hat\sigma_{gg}\over d\ktat d\ktbt}\ =\
\biggl[{C_A\alpha_s\over \ktat}\biggr] \,
{\bar f}(\qta,\qtb,\dy) \,
\biggl[{C_A\alpha_s\over \ktbt}\biggr] \ ,
\label{cross}
\eeq
where $C_{A}=N_{c}=3$, and $\qta$, $\qtb$ are the 
momenta transferred in the $t$-channel, with $\vec{q}_{a\perp}=-\vkta$ 
and $\vec{q}_{b\perp}=\vktb$. The quantities in square brackets
are the LO impact factors for jet production. 
The function ${\bar f}(\qta,\qtb, \dy)$ is the
azimuthally-averaged Green's function associated with the gluon exchanged 
in the crossed channel, and  is obtained by integrating out the
azimuthal angles in the solution of the BFKL equation,
\beq
{\bar f}(\qta,\qtb,\dy)\, = {1\over 4 \sqrt{\qtat \qtbt}} 
\int_{-\infty}^{\infty} d\nu\, 
e^{\omega(\nu)\dy}\, \left(\qtat\over \qtbt
\right)^{i\nu}\, ,
\label{solc}
\eeq
with
\beqa
\omega(\nu) &=& - 2{\bar \alpha_s}\, \left[ {\rm Re}\, \psi\left(1/2 
+i\nu\right) - \psi(1)\right] \label{om}
\\* 
&=& A - B\nu^2 + {\cal O}(\nu^4)
\, ,
\label{omexp}
\eeqa
with $\psi$ the digamma function, and 
\beq
\bar\alpha_s\,=\,\alpha_s N_c/\pi,\;\;\;\;
A\,=\,4\bar\alpha_s\ln{2},\;\;\;\;
B\,=\,14\zeta(3)\,\bar\alpha_s,
\label{asAB}
\eeq

\subsection{The standard Mueller-Navelet analysis}
\label{sec:twoone}

In order to elucidate how the D0 Collaboration~\cite{Abbott:2000ai}
evaluates the effective BFKL intercept, we follow the original Mueller-Navelet 
approach~\cite{Mueller:1987ey}, i.e. we substitute \eqns{cross}{solc} into
Eq.~(\ref{mrfac}) and integrate it over the transverse momenta $\kta$ and 
$\ktb$ above
a threshold $\Ecut$, at fixed coupling $\alpha_s$ and fixed $x_a^0,\, x_b^0$. 
The rapidity interval $\dy = |y_a-y_b|$ in \eqn{solc} is determined from the
$x$'s (\ref{nkin0}),
\beq
\dy = \ln{x_a^0 x_b^0 S\over \kta\ktb} \label{rapint}
\eeq
and since it depends on $\kta\ktb$, it is not a constant 
within the integral. However, the dominant, i.e. the leading
logarithmic, contribution to \eqn{solc} comes from the largest value
of $\dy$, which is attained at the transverse momentum threshold, thus
in Ref.~\cite{Mueller:1987ey} $\dy$ is fixed at its maximum by reconstructing 
the $x$'s at the kinematic threshold for jet production and setting them in a 
one-to-one correspondence with the jet rapidities
\beq
x_a^\mn = \frac{\Ecut}{\sqrt{S}} e^{y_a}\qquad \qquad
x_b^\mn = \frac{\Ecut}{\sqrt{S}} e^{-y_b}\, .
\label{mnkin}
\eeq
Then the factorization formula (\ref{mrfac}) is determined at fixed 
$x_a^\mn, x_b^\mn$.
Having fixed the rapidity interval (\ref{rapint}) to
\beq
\dy = \ln{x_a^\mn x_b^\mn S\over \Ecutt} \, ,\label{mnrap}
\eeq
the integration over $\kta$ and $\ktb$ can be 
straightforwardly performed\footnote{In order to do the integrals analytically,
it is necessary to fix the factorization scale $\mu_F$ in Eq.~(\ref{mrfac}), 
{\sl e.g.} $\mu_F = \Ecut$.}, and
the gluon-gluon cross section~(\ref{cross}) becomes
\beq
\hat\sigma_{gg}(\kta\!>\!\Ecut,\ktb\!>\!\Ecut)\, = 
{\pi C_A^2\alpha_s^2\over 2 \Ecutt} \cF(\dy,1)\, ,
\label{kintsol}
\eeq
with
\beq
\cF(z,t) = {1\over 2\pi} \int_{-\infty}^{\infty} d\nu\, 
{e^{\omega(\nu)z}\over \nu^2 + 1/4}\, \cos(2\nu\ln t)\, .\label{ffunct}
\eeq
For $\bar \alpha_s \dy \ll 1$, we can expand \eqn{ffunct} and 
obtain~\cite{Mueller:1987ey}
\beq
\hat\sigma_{gg}(\kta\!>\!\Ecut,\ktb\!>\!\Ecut)\, = 
{\pi C_A^2\alpha_s^2\over 2 \Ecutt}\left[ 1 +
{\cal O}\left( (\bar\alpha_s\dy)^2 \right) \right]\, .
\label{ysmallsol}
\eeq
On the other hand, for $\dy\gg 1$ we can perform a saddle-point evaluation 
of Eq.~(\ref{ffunct}), and using the small-$\nu$ expansion (\ref{omexp}), 
we obtain the asymptotic behaviour of the gluon-gluon cross 
section~\cite{Kuraev:1977fs,Balitsky:1978ic}
\beq
\hat\sigma_{gg}^{(\dy\gg 1)}(\kta\!>\!\Ecut,\ktb\!>\!\Ecut)\, = 
{\pi C_A^2\alpha_s^2\over 2 \Ecutt}
{e^{A \dy}\over \sqrt{\pi B \dy/4}}\, .\label{asympsol}
\eeq
At very large rapidities the resummed gluon-gluon cross section grows
exponentially with $\dy$, in contrast to the LO (${\cal O}(\alpha_s^2)$)
cross section~(\ref{ysmallsol}), which is constant at large $\dy$. From the
asymptotic formula~(\ref{asympsol}) the effective BFKL intercept
\mbox{$\alpha_{\sss BFKL}\equiv A+1$} can be derived.  In the experiment of
Ref.~\cite{Abbott:2000ai} the BFKL intercept is measured by considering the
ratio of hadronic cross sections, Eq.~(\ref{mrfac}), obtained at different
centre-of-mass energies and at fixed $x_{1,2}$ and scale. This allows the
dependence on the parton densities to cancel, and the ratio of hadronic cross
sections is therefore approximately equal to that of partonic cross sections
evaluated at the relevant $\dy$ values.

In \eqnss{kintsol}{asympsol} we have summarised the standard Mueller-Navelet
analysis in which it is assumed that the $x$'s are reconstructed through
\eqn{mnkin}, and that the jet transverse momenta are unbounded from above.
However, this is not the case for the D0 analysis, since 
\begin{itemize}
\item[$a)$] D0 collect data with an upper bound on $Q^2=\kta\ktb$, which is
of the same order of magnitude as the square of the lower cut on the jet 
transverse momenta, and thus cannot be ignored in the integration
over the transverse momenta.
\item[$b)$]D0 reconstruct the $x$'s through \eqn{eq:one}, which is well
approximated by \eqn{nkin0}, but not by \eqn{mnkin};
\end{itemize}
We examine these two issues, and the modifications they entail on
\eqnss{kintsol}{asympsol}, in turn.

\subsection{Dijet production with an upper bound on $Q^2$}
\label{sec:twotre}
In the Mueller-Navelet analysis the integration
over the transverse momenta is taken up to infinity on the grounds
that a finite and large upper bound on the transverse momenta would entail
a contribution which is power suppressed in the ratio of the jet threshold to
the upper bound.
However, D0 collect data with an upper bound on $Q^2=\kta\ktb$, 
namely $Q^2_{\rm max}$ = 1000 ${\rm GeV}^2$, while 
$Q^2_{\rm min}=\Ecutt$ = 400 ${\rm GeV}^2$. 
When $Q^2_{\rm min} \sim Q^2_{\rm max}$, the upper bound cannot 
be ignored in the integration over the transverse momenta. 

\begin{figure}[htb]
\centerline{
        \epsfig{file=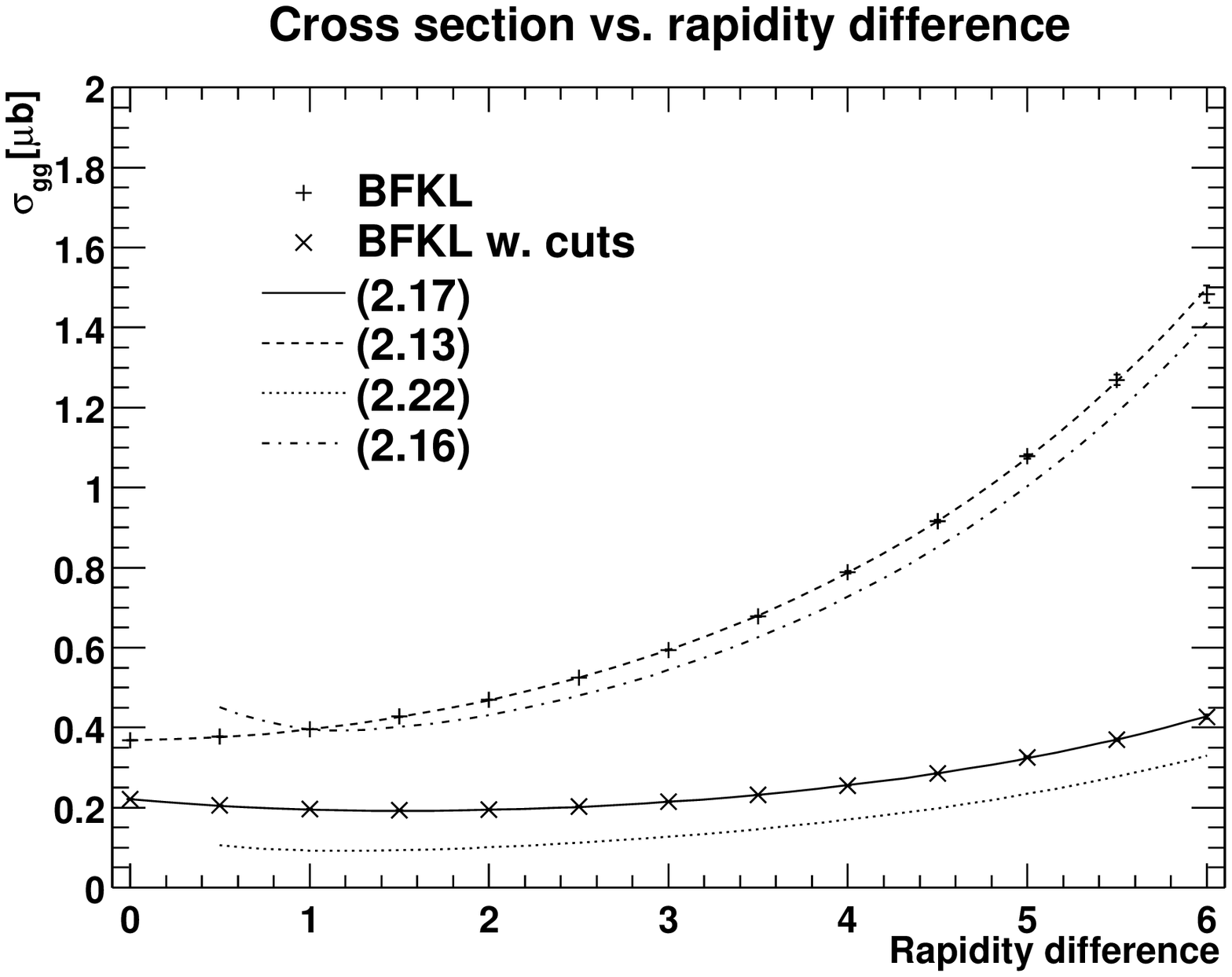,width=13cm} }
\ccaption{}{ \label{ggxsec}
The dependence of the  LL BFKL gluon-gluon subprocess cross section 
on the dijet rapidity separation $\Delta y$, without
(Eq.~(\ref{kintsol}),  upper dashed line)
and with (Eq.~(\ref{qmaxnaive}), lower solid line) the $Q^2_{\rm max}$
cut. The `data points' are the same quantities calculated using
the BFKL Monte Carlo discussed in Section~\protect\ref{sec:MCBFKL},
with $\alpha_s$ fixed and no additional kinematic cuts. Also shown
are the asymptotic $\dy\gg 1$ approximations, Eqs.~(\ref{asympsol}) and (\ref{asympmax}).
The parameter 
values are $\alpha_s = 0.164$, $ \Ecut = 20$~GeV, 
$Q^2_{\rm max} = 1000$~GeV$^2$.
}
\end{figure}

In order to assess what the modification on \eqnss{kintsol}{asympsol} is,
we integrate
the gluon-gluon cross section~(\ref{cross}) over the transverse momenta 
$\kta$ and $\ktb$ above a threshold $\Ecut$ with the
upper cut \mbox{$Q^2<Q^2_{\rm max}$} imposed. We obtain
\beq
\hat\sigma_{gg}(\kta\!>\!\Ecut,\ktb\!>\!\Ecut,
\kta\ktb\!\!< Q^2_{\rm max})= 
{\pi C_A^2\alpha_s^2\over 2 \Ecutt} 
\left[ \cF(\dy,1) - {\Ecutt\over Q^2_{\rm max}} 
\cG\left(\dy,{\Ecutt\over Q^2_{\rm max}} \right)\right]\, ,
\label{qmaxnaive}
\eeq
with rapidity interval $\dy$ defined in \eqn{mnrap}, $\cF$ defined 
in \eqn{ffunct}, and 
\beq
\cG(z,t) = \cF(z,t) - {1\over 2\pi} \int_{-\infty}^{\infty} d\nu\, 
{e^{\omega(\nu)z}\over \nu^2 + 1/4} 
{\sin\left(2\nu\ln{t} \right)\over 2\nu}\, .\label{gfunct}
\eeq
The analytic form of \eqn{qmaxnaive} depends on the particular definition
of the upper cutoff $Q^2_{\rm max}$ that D0 uses, and changes 
substantially the shape of the gluon-gluon cross section (see \fig{ggxsec})
and in particular its subasymptotic dependence on 
$\Delta y$.
At $\bar \alpha_s \dy \ll 1$, we expand the exponentials in \eqn{qmaxnaive}
and obtain 
\beq
\hat\sigma_{gg}(\kta\!>\!\Ecut,\ktb\!>\!\Ecut,
\kta\ktb\!\!< Q^2_{\rm max})\, 
= {\pi C_A^2\alpha_s^2\over 2 \Ecutt}\left[ 1 - 
{\Ecutt\over Q^2_{\rm max}} +
{\cal O}\left( \bar\alpha_s\dy \right) \right]\, .\label{ysmallqmax}
\eeq
Thus for the D0 cuts, $Q^2_{\rm max}$ = 1000 ${\rm GeV}^2$, 
\eqn{ysmallqmax} lowers the LO cross section 
(\ref{ysmallsol}) by 40\%.
At $\dy\gg 1$ a saddle-point evaluation of \eqn{qmaxnaive} yields
\beqa
\lefteqn{
\hat\sigma_{gg}^{(\dy\gg 1)}(\kta\!>\!\Ecut,\ktb\!>\!\Ecut,
\kta\ktb\!\!< Q^2_{\rm max})\
=\ {\pi C_A^2\alpha_s^2\over 2 \Ecutt}\ e^{A \dy} } \label{qmaxasympt}
\\*
&\times& \Biggl\{ e^{B\dy/4}
\left[ \Phi\left({\ln(Q^2_{\rm max}/\Ecutt)+B\dy/2\over \sqrt{B\dy}}
\right) - \Phi\left({\sqrt{B\dy}\over 2}\right) \right] - 
{\Ecutt\over Q^2_{\rm max}} \Phi\left({\ln(Q^2_{\rm max}/\Ecutt)
\over \sqrt{B\dy}} \right) \Biggr\}\, ,\nonumber
\eeqa
with the error function
\beq
\Phi(x) = {2\over\sqrt{\pi}} \int_0^x dt e^{-t^2}\, .\label{error}
\eeq
Using the asymptotic expansion of the error function at $x\gg$ 1, for
very large rapidities $\sqrt{B\dy}\gg$ 1 and at fixed 
$Q^2_{\rm max}/\Ecutt$ we obtain
\beq
\hat\sigma_{gg}^{(\dy\gg 1)}(\kta\!>\!\Ecut,\ktb\!>\!\Ecut,
\kta\ktb\!\!< Q^2_{\rm max})\, 
= {\pi C_A^2\alpha_s^2\over 2 \Ecutt}
{e^{A \dy}\over \sqrt{\pi B \dy/4}} \left[1 - {\Ecutt\over Q^2_{\rm max}}
\left( 1 + \ln{Q^2_{\rm max}\over \Ecutt}\right) \right],\label{asympmax}
\eeq
which is simply the asymptotic cross section of Eq.~(\ref{asympsol}) reduced
by a constant factor. For the D0 values of $\Ecutt$ and $Q^2_{\rm max}$, this
corresponds to a reduction by a factor of about 4.3 in the standard asymptotic 
formula (\ref{asympsol}).

Although it might appear from Eq.~(\ref{asympmax}) 
that the only effect of the $Q^2_{\rm max}$ cut is to change the 
normalization relative to Eq.~(\ref{asympsol}), which would drop out
of the ratio of cross sections, 
one has to keep in mind that both equations are derived (from 
Eqs.~(\ref{qmaxnaive}) and~(\ref{kintsol}) respectively) in the 
asymptotic limit $\dy\gg 1$. In Fig.~\ref{ggxsec} we plot both the
integral formulae and their
asymptotic solutions.  We see that the differences 
\mbox{$\hat\sigma_{gg}-\hat\sigma_{gg}^{(\dy\gg 1)}$} 
are roughly
constant with respect to $\dy$, and thus the relative differences
get smaller with increasing $\dy$. However, at the $\dy$ values
relevant to D0 analysis, it appears that non-negligible 
subleading corrections to the asymptotic formulae
should be taken into account when determining the effective BFKL 
intercept. As can be inferred from Fig.~\ref{ggxsec}, these effects
are more important when a $Q^2$ cut is imposed, since in this
case it takes longer for the exponential rise with $\dy$ to set in.

In conclusion, the effect of an upper bound on the product of the jet
transverse momenta can have a significant effect on both the
normalization and $\Delta y$ dependence of the gluon-gluon cross section.
For the D0 values, the increase of the cross section from small
to large $\Delta y$ is weakened by a factor of approximately 2, as shown
in \fig{ggxsec}.
Care must therefore be taken in attributing any observed cross section
increase exclusively to the $e^{A \dy}$ `BFKL' factor.

\subsection{Dijet production at $x$'s fixed as in the D0 set-up}
\label{sec:twotwo}

In the analysis performed by D0, the $x$'s are reconstructed through \eqn{eq:one}.
Since the jets are selected by requiring that $\dy > 2$, \eqn{nkin0} is a good 
approximation to \eqn{eq:one}. Conversely, the $x$'s (\ref{mnkin}) 
used in the Mueller-Navelet analysis are by definition a good approximation 
to the D0 $x$'s (\ref{eq:one}) only at the kinematic threshold for jet 
production. Therefore in this section we shall examine
the modifications induced on \eqnss{kintsol}{asympsol} by defining
the $x$'s as in  \eqn{nkin0}. First, we note that in this case the jet 
rapidities are not fixed, rather in a given ($x_a^0,\,x_b^0$) bin all the 
transverse momenta and rapidities contribute which fulfil \eqn{nkin0}.
Thus the rapidity interval between the jets cannot be used as an independent,
fixed observable.
For convenience, we rewrite the rapidity interval (\ref{rapint}) as 
\beq
\dy = Y + \ln{\Ecutt\over \kta\ktb}\, ,\label{uno}
\eeq
with\footnote{The constant $Y$ resembles the rapidity interval (\ref{mnrap})
used in the Mueller-Navelet analysis, however it is not the same since
\eqn{uno} entails that $\dy\le Y$, while \eqns{rapint}{mnrap} are just two 
different ways of defining the same rapidity interval.}
\beq
Y = \ln{x_a^0x_b^0 S\over\Ecutt}\, .\label{const}
\eeq
The requirement that the rapidity interval be positive, $\dy\ge 0$,
imposes an effective upper bound on $Q^2$, 
\beq
Q^2_{\rm max} = \Ecutt e^Y\, .\label{qmax1} 
\eeq
Integrating then the
gluon-gluon cross section~(\ref{cross}) over $\kta$ and $\ktb$ above 
$\Ecut$, at fixed $x_a^0,\, x_b^0$ and fixed coupling $\alpha_s$,
we obtain
\beq
\hat\sigma_{gg}(\kta\!>\!\Ecut,\ktb\!>\!\Ecut,
\kta\ktb\!\!< Q^2_{\rm max})= 
{\pi C_A^2\alpha_s^2\over 2 \Ecutt} 
\left[ \cFt(Y,1) - {\Ecutt\over Q^2_{\rm max}} 
\cGt\left(Y,{\Ecutt\over Q^2_{\rm max}} \right)\right]\, ,
\label{qmaxcomb}
\eeq
with
\beqa
\cFt(z,t) &=& {1\over 2\pi} \int_{-\infty}^{\infty} d\nu\, 
{e^{\omega(\nu)z}\over \nu^2 + \displaystyle{[1+\omega(\nu)]^2\over 4}}\, 
t^{\omega(\nu)}\, \cos(2\nu\ln t)\, ,
\label{ftfunct}
\\
\cGt(z,t) &=& \cFt(z,t) - {1\over 2\pi} \int_{-\infty}^{\infty} d\nu\, 
{e^{\omega(\nu)z}\over \nu^2 + \displaystyle{[1+\omega(\nu)]^2\over 4}}\,
t^{\omega(\nu)} (1+\omega(\nu))\frac{\sin\left(2\nu\ln{t}\right)}{2\nu}\, ,
\label{gtfunct}
\eeqa
and $\omega(\nu)$ as in \eqn{om}. Note that as $Y\to 0$ in \eqn{qmax1}, 
the upper bound on $Q^2$ goes to the kinematic threshold, 
$Q^2_{\rm max} \to\Ecutt$, and accordingly the cross section (\ref{qmaxcomb})
vanishes. Note also that the tilde functions $\cFt$~(\ref{ftfunct}) and 
$\cGt$~(\ref{gtfunct}) reduce to the functions $\cF$~(\ref{ffunct})
and $\cG$~(\ref{gfunct}) for $\alpha_s\ll 1$ but $\alpha_s Y\approx1$.  
This is understandable because the limit $\alpha_s\ll 1$ is equivalent 
to neglecting subleading corrections, since $(\alpha_s)^2\Delta y \ll 
\alpha_s \Delta y$, and therefore to neglecting differences in the 
definition of the rapidity interval, $\Delta y\approx Y$.
For $Y\gg 1$ we perform a saddle-point evaluation of \eqn{qmaxcomb},
and obtain the asymptotic behaviour
\beqa
\lefteqn{\hat\sigma_{gg}^{(Y\gg 1)}(\kta\!>\!\Ecut,\ktb\!>\!\Ecut,
\kta\ktb\!\!< Q^2_{\rm max})\, =}\nonumber\\ 
&& {\pi C_A^2\alpha_s^2\over 2 \Ecutt}
{e^{A Y}\over \sqrt{\pi B Y/4}} {1\over (1+A)^2}
\left[1 - \left( {\Ecutt\over Q^2_{\rm max}}
\right)^{1+A} \left( 1 + (1+A)\ln{Q^2_{\rm max}\over \Ecutt}\right) \right]\,
.\label{asympcombmax}
\eeqa

\begin{figure}[htb]
  \centerline{ \epsfig{file=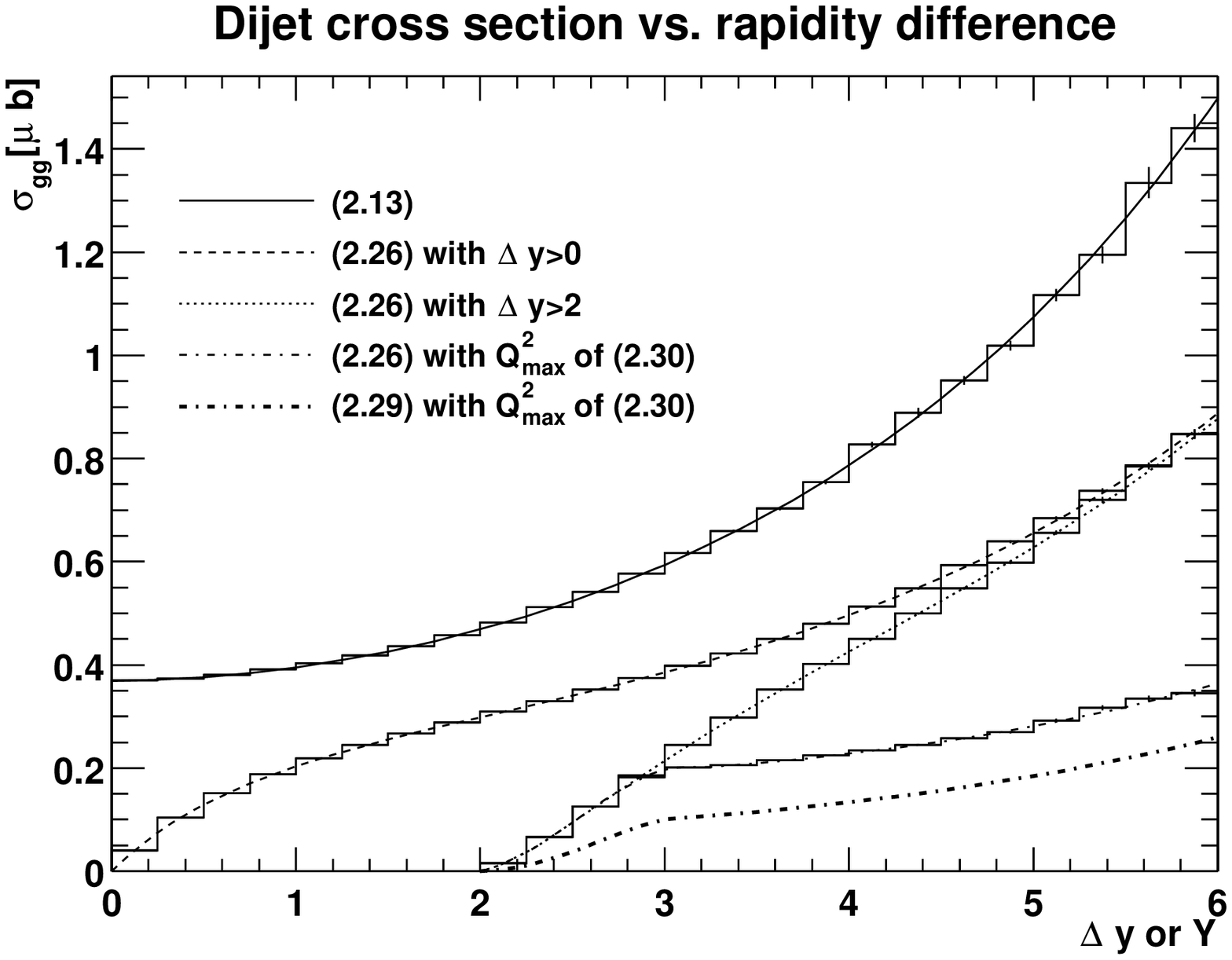,width=13cm} } \ccaption{}{
    \label{d0ggxsec} The dependence of the LL BFKL gluon-gluon cross section
    on $\Delta y$ in the standard Mueller-Navelet calculation
    (Eq.~(\ref{kintsol})) (upper solid line) and on $Y$ for the D0 setup
    (Eq.~(\ref{qmaxcomb})). Four curves are shown for the definition of $x$'s
    applied in the D0 analysis: Dashed line for the requirement $\Delta y>0$,
    dotted line for $\Delta y>2$, dash-dotted for $Q^2_{\rm max}$ of
    Eq.~(\ref{qmax2}) and finally the lower, fat dash-dotted line for the
    asymptotic behaviour (Eq.~(\ref{asympcombmax})) using $Q^2_{\rm max}$ of
    Eq.~(\ref{qmax2}). The histograms are filled using the MC.

}
\end{figure}

We can also use the above analysis to include the D0 experimental cuts
of $Q^2<1000$ GeV$^2$ and $\Delta y>2$.
In this case the analysis holds unchanged except that the upper bound
on $Q^2$ is given by
\beq
Q^2_{\rm max} = {\rm min}(1000\, {\rm GeV}^2,\,\Ecutt e^{(Y-2)})\, ,\label{qmax2}
\eeq
where we have used the fact that $\Delta y>2$ imposes the second
effective upper bound on $Q^2$.
The shape of the cross section as a function of $Y$ depends 
crucially on whether the upper bound on $Q^2$ is given by \eqn{qmax1} or 
(\ref{qmax2}) (see \fig{d0ggxsec}). This is more clearly apparent
in the asymptotic region, $Y\gg 1$, since for the upper bound (\ref{qmax1})
we can safely take $Q^2_{\rm max}\to\infty$, with only the first term
in the square brackets of \eqn{asympcombmax} contributing; conversely,
when the upper bound is given by \eqn{qmax2}, the sharp cutoff 
$Q^2_{\rm max}$ = 1000 ${\rm GeV}^2$ is much more restrictive than
the bound (\ref{qmax1}) and depletes the cross section, which is 
given by the whole \eqn{asympcombmax}. 

Using \fig{d0ggxsec}, we can get some idea of the expected effect of the
$\Delta y>2$ cut on the cross section ratio measured by D0. {}From
\eqn{qmax2} we see that this cut is inconsequential when
\beq
Y>2+\ln(1000\ \rm{GeV}^2/\Ecutt)\simeq2.92\, ,\label{dycut}
\eeq
where we have used $E_\perp=20$ GeV. 
Conversely, this cut removes the entire cross section for $Y\le2$.
{}For $\sqrt{S}=1800$ GeV we find $Y>2.92$
for all bins, so the cut has no effect.  However,
for $\sqrt{S}=630$ GeV we find $Y=2.37$ in bin 1, $Y=2.63$ in bin
2, and $Y=2.88$ in bin 3, where we have used the minimum $x_1$
and $x_2$ in each bin to evaluate $Y$.  Thus, bins 1 and 2 (and to
some extent bin 3) get depleted at 630 GeV, simply due to the
$\Delta y>2$ cut.  In section 4 we will see that this leads to
a large cross section ratio in these bins, independent of the BFKL
dynamics.

Finally, we note that the asymptotic cross section
Eq.~(\ref{asympcombmax}) has the same shape in $Y$ as Eq.~(\ref{asympsol}) in
$\dy$ but different normalization: at $\alpha_s(Q^2=400 {\rm GeV}^2) = 0.164$, 
the normalization of
Eq.~(\ref{asympcombmax}) with upper bound (\ref{qmax1}) is a factor 2.1
smaller than the one of the standard asymptotic formula (\ref{asympsol}), 
which becomes a factor 5.4 smaller than the one of
Eq.~(\ref{asympsol}) if the upper bound (\ref{qmax2}) is used.
However, as \fig{d0ggxsec} shows, for the values of rapidity interval relevant 
to the D0 analysis we are far from the asymptotic region, and thus all the
caveats made at the end of \sec{sec:twotre} on the extraction of the BFKL
intercept from the D0 data apply in this case as well.

\section{Dijet production and energy-momentum conservation}
\label{sec:MCBFKL}

Besides the effects described in the previous section, there are other 
facts that need to be considered when aiming at comparing experimental 
data to BFKL predictions, some of general nature and some specific to the
experimental set-up of Ref.~\cite{Abbott:2000ai}. We will comment 
on the latter in the next section. As far as the former are
concerned, we remind the reader of the following facts.  $(i)$ The LL BFKL
resummation is performed at fixed coupling constant, and thus any variation
in the scale at which $\alpha_s$ is evaluated appears in the
next-to-leading-logarithmic (NLL) terms.  $(ii)$ Because of the strong
rapidity ordering, any two-parton invariant mass is large. Thus there are no
collinear divergences in the LL BFKL resummation; jets are determined only at
LO and accordingly have a trivial structure.  $(iii)$ Energy and longitudinal
momentum are not conserved, and since the momentum fractions $x$ of the
incoming partons are reconstructed from the kinematic variables of the
outgoing partons, the BFKL theory can severely underestimate the exact value
of the $x$'s, and thus grossly overestimate the parton luminosities.  In
fact, if $n+2$ partons are produced, energy-momentum conservation gives
\beqa
x_a &=&   \frac{e^{y_a}}{\sqrt{S}}
   \left( \kta \,     +\,  \ktb e^{-\dy} \,   
+ \, \sum_{i=1}^n k_{i\perp} e^{y_i-y_a } \right) \, ,  
\nonumber \\
x_b &=&   \frac{e^{-y_b}}{\sqrt{S}}
   \left(  \ktb  \,  + \,          \kta e^{-\dy} \,  
+ \, \sum_{i=1}^n k_{i\perp} e^{-y_i+y_b} \right) \, .
\label{emcons}
\eeqa
The momentum fractions (\ref{nkin0}) in the high-energy limit
are recovered by imposing the strong rapidity ordering (\ref{mrk}). 
However, the requirement $x_a,x_b \leq 1$ effectively imposes
an upper limit on the transverse momentum ($\vec k_{i\perp}$) integrals.
This effect is completely analogous to that we have considered in 
sections~\ref{sec:twotre} and~\ref{sec:twotwo}.
 
In an attempt to go beyond the analytic leading-logarithm BFKL results,
a Monte Carlo approach has been 
adopted~\cite{Schmidt:1997fg,Orr:1997im,Orr:1998hc}. If we transform 
the relevant Green's function to moment space via
\begin{equation}
f(\vec q_{a\perp},\vec q_{b\perp},\Delta y) \ =\ \int {d\omega\over 2\pi i}\, 
e^{\omega\Delta y}\, 
f_{\omega}(\vec q_{a\perp},\vec q_{b\perp})\,,\label{moment}
\end{equation}
(by averaging $f$ over azimuthal angles one obtains the quantity
$\bar{f}$ introduced in Eq.~(\ref{cross})) we can write the BFKL 
equation as
\begin{equation}
\omega\, f_{\omega}(\vec q_{a\perp},\vec q_{b\perp})\, =
{1\over 2}\,\delta (\vec q_{a\perp}-\vec q_{b\perp})\, +\, 
{\bar \alpha_s \over \pi} 
\int {d^2\vec k_{\perp}\over k_{\perp}^2}\,
K(\vec q_{a\perp},\vec q_{b\perp},\vec k_{\perp})\, ,\label{bfklb}
\end{equation}
where the kernel $K$ is given by
\begin{equation}
K(\vec q_{a\perp},\vec q_{b\perp},\vec k_{\perp}) = f_{\omega}(\vec
q_{a\perp}+\vec k_{\perp},
\vec q_{b\perp}) - {q_{a\perp}^2\over k_{\perp}^2 + 
(\vec q_{a\perp}+\vec k_{\perp})^2}\, f_{\omega}(\vec q_{a\perp},\vec q_{b\perp})
\, .\label{kern}
\end{equation}
The first term in the kernel accounts for the emission of a real gluon of
transverse momentum $\vec k_{\perp}$ and the second term accounts for the
virtual radiative corrections. By solving the BFKL equation (\ref{bfklb}) by
iteration, which amounts to `unfolding' the summation over the intermediate
radiated gluons and making their contributions explicit, it is possible to
include the effects of both the running coupling and the overall kinematic
constraints.  It is also straightforward to implement the resulting iterated
solution in an event generator.

The first step in this procedure is to separate the $\vec k_{\perp}$ integral
in (\ref{bfklb}) into `resolved' and `unresolved' contributions, according to
whether they lie above or below a small transverse energy scale $\mu$.  The
scale $\mu$ is assumed to be small compared to the other relevant scales in
the problem (the minimum transverse momentum $\Ecut$ for example).  The
virtual and unresolved contributions are then combined into a single, finite
integral.  The BFKL equation becomes
\begin{eqnarray}
\omega\, f_{\omega}(\vec q_{a\perp},\vec q_{b\perp}) & =&
{1\over 2}\,\delta (\vec q_{a\perp}-\vec q_{b\perp})\, 
+\, {\bar\alpha_s\over \pi} 
\int_{k_{\perp}^2 > \mu^2} {d^2\vec k_{\perp}\over k_{\perp}^2}\,
f_{\omega}(\vec q_{a\perp}+\vec k_{\perp},\vec q_{b\perp}) \nonumber \\
&+&  {\bar\alpha_s\over \pi} \int {d^2\vec k_{\perp}\over k_{\perp}^2}
\left[  f_{\omega}(\vec q_{a\perp}+\vec k_{\perp},\vec q_{b\perp})\, \theta(\mu^2 -k_{\perp}^2)\, - \, {q_{a\perp}^2 \, f_{\omega}(\vec q_{a\perp},\vec q_{b\perp}) 
 \over k_{\perp}^2 + (\vec q_{a\perp}+\vec k_{\perp})^2}
\right] .\label{bfklbx}
\end{eqnarray}
The combined unresolved/virtual integral can be simplified by noting 
that since $ k_{\perp}^2 \ll q_{a\perp}^2,q_{b\perp}^2$ by construction,
the $\vec k_{\perp}$ term in the argument of $f_{\omega}$ can be neglected,
giving
\begin{equation}
(\omega - \omega_0)\, f_{\omega}(\vec q_{a\perp},\vec q_{b\perp}) \, =\,
{1\over 2}\,\delta (\vec q_{a\perp}-\vec q_{b\perp})\, 
+\, {\bar\alpha_s\over \pi} 
\int_{k_{\perp}^2 > \mu^2} {d^2\vec k_{\perp}\over k_{\perp}^2}\,
f_{\omega}(\vec q_{a\perp}+\vec k_{\perp},\vec q_{b\perp}) \nonumber \\
\, ,\label{bfklcx}
\end{equation}
where
\begin{equation}
\omega_0  = {\bar\alpha_s\over \pi} 
\int {d^2\vec k_{\perp}\over k_{\perp}^2}
\left[   \theta(\mu^2 -k_{\perp}^2)\, - \, {q_{a\perp}^2 
 \over \vec k_{\perp}^2 + (\vec q_{a\perp}+\vec k_{\perp})^2}
\right] 
\simeq {\bar\alpha_s}\, \ln\left( { \mu^2 \over q_{a\perp}^2 }   \right) 
\, .    \label{eq:omega0}
\end{equation}
The virtual and unresolved contributions are now contained
in $\omega_0$ and we are left with an integral over resolved real
gluons. We can now solve (\ref{bfklcx}) iteratively, 
and performing the inverse transform
we have 
\begin{equation}
f(\vec q_{a\perp},\vec q_{b\perp},\Delta y)\, =  \, 
  \sum_{n=0}^{\infty} f^{(n)}(\vec q_{a\perp},\vec q_{b\perp},\Delta y) \; .
\label{eq:b7}
\end{equation}
where
\begin{eqnarray}
f^{(0)}(\vec q_{a\perp},\vec q_{b\perp},\Delta y) &  =  &  
\left[ \frac{\mu^2}{q_{a\perp}^2} \right]^{\bar\alpha_s\Delta y}\,
\,\frac{1}{2}\, \delta (\vec q_{a\perp}-\vec q_{b\perp} )\,,
\nonumber \\
f^{(n\geq 1)}(\vec q_{a\perp},\vec q_{b\perp},\Delta y) &  =  &  
\left[ \frac{\mu^2}{q_{a\perp}^2} \right]^{\bar\alpha_s\Delta y}\,
\left\{ \prod_{i=1}^{n}  \int d^2 \vec k_{i\perp}\, dy_i \, {\cal F}_i \right\}
\,\frac{1}{2}\, \delta (\vec q_{a\perp}-\vec q_{b\perp} - 
\sum_{i=1}^n \vec k_{i\perp})\,,
\nonumber \\
{\cal F}_i &=& \frac{\bar\alpha_s}{\pi k_{i\perp}^2}\, 
\theta(k_{i\perp}^2 -\mu^2)\, \theta(y_{i-1}-y_i)\,
\left[ { (\vec q_{a\perp} +\sum_{j=1}^{i-1}\vec k_{j\perp}  )^2
 \over (\vec q_{a\perp} +\sum_{j=1}^{i}\vec k_{j\perp}  )^2 }
\right]^{\bar\alpha_s y_i}\,.
\label{eq:b8}
\end{eqnarray}
Thus the solution to the BFKL equation is recast in terms of phase space
integrals for resolved gluon emissions, with form factors representing the
net effect of unresolved and virtual emissions. In this way, each $f^{(n)}$
depends on the resolution parameter $\mu$, whereas the full sum $f$ does not.
Changing $\mu$ simply shifts parts of the cross section from one $f^{(n)}$ to
another.\footnote{One other technical point deserves comment. The
  decomposition of $f$ into the individual $f^{(n)}$ breaks the symmetry
  between $\vec q_{a\perp}$ and $\vec q_{b\perp}$ which is only restored in
  the sum over $n$. Care must be taken when imposing {\it asymmetric} cuts on
  the $a$ and $b$ jets (see following section).  Numerically, it is found
  that the $\mu$ dependence is slightly weaker when the higher $\Ecut$ cut is
  imposed on jet $a$. This asymmetry between the two jets is an artifact of
  the recursive solution.}

  Unlike in the case of DGLAP evolution, there is no strong
ordering of the transverse momenta $k_{i\perp}$.  Strictly speaking, the
derivation given above only applies for fixed coupling because we have left
$\alpha_s$ outside the integrals.  The modifications necessary to account for
a running coupling $\alpha_s (k_{i\perp}^2)$ are
straightforward~\cite{Orr:1997im}.

The expression for $f$ in Eqs.~(\ref{eq:b7}) and (\ref{eq:b8})  is amenable to
numerical integration which can be  implemented in the form of 
a Monte Carlo event generator~\cite{Schmidt:1997fg,Orr:1997im,Orr:1998hc}. 
In this way we can trivially impose further cuts and combine the subprocess
cross section with parton densities. Having made explicit the BFKL gluon
emission phase space,  overall energy and momentum conservation
is imposed by using the momentum fraction definitions in Eq.~(\ref{emcons}).
 We can also
run the Monte Carlo in `LL mode', i.e. with fixed $\alpha_s$ and no 
additional cuts or parton densities, 
to check that it does indeed reproduce the analytic results
of the previous section, see~\fig{ggxsec}. 

\section{Equal transverse momentum cuts: a dangerous choice}
\label{sec:equalcut}

We now take a closer look at the set-up specific to the D0 analysis
of Ref.~\cite{Abbott:2000ai}. As a preliminary observation, we might say 
that the values $\dy$ probed are quite far from the asymptotic region where 
\eqn{asympsol} is expected to hold, particularly at $\sqrt{S}$ = 
630~GeV, where $\dy$ is of the order of 2 to 3; unfortunately, here the
only solution is to wait for the LHC to come into operation. A more serious, 
but solvable, problem is the following: dijet rates are quite sensitive to 
the emission of soft and collinear gluons, in the case in which they are
defined by imposing equal cuts on the transverse energies of the two
tagged jets.  In this sense, a dijet total rate is completely analogous to
the azimuthal correlation mentioned above. A detailed discussion on
this point is given in Ref.~\cite{Frixione:1997ks}, and  will not be 
repeated here. In the current study, we will limit ourselves to
illustrating the discussion of Ref.~\cite{Frixione:1997ks} by
means of examples relevant to dijet production at the Tevatron.
We will do this in two steps. First, in Subsection~\ref{sec:FO},
we will study this issue using a fixed-order perturbative computation, 
showing that dijet cross sections defined with unequal transverse 
momentum cuts do not have the same problems as those defined with 
equal cuts. Then, in Subsection~\ref{sec:BFKLcuts}, we will repeat 
the analysis of Section~\ref{sec:naiveBFKL} in the more general case
of unequal transverse momentum cuts. 

\subsection{Dijet cross sections at fixed perturbative order}
\label{sec:FO}

In this subsection\footnote{In the fixed order pQCD analysis, at variance with
  Section~\ref{sec:naiveBFKL}, we set $\muf=(\kta +\ktb)/2$, a choice more
  suited to a Monte Carlo approach.}, we consider jet production at fixed
perturbative order in QCD. In particular, we use the partonic event generator
of Ref.~\cite{Frixione:1997np}, which is accurate to NLO for any one- or
two-jet observables. Similarly to what has been done previously for the
gluon-gluon cross section, we define a total dijet cross section as follows:
\beq \sigma(\de,\Cuts)=\sigma(\kta\!>\!\Ecut,\ktb\!>\!\Ecut+\de,\Cuts),
\label{ratedef}
\eeq
where $\Cuts$ generically indicates a set of cuts to be added to
transverse-momentum cuts. As already mentioned, D0 have
\beq
\Cuts:\;\;\;\;\;\;
\abs{y_i}<3,\;\;
\dy>2,\;\;
Q^2<1000~{\rm GeV}^2
\label{Cdef}
\eeq
($i=a,b$), together with some additional cuts on $x_1$ and $x_2$; furthermore,
$\Ecut=20$~GeV.

\begin{figure}
\centerline{
   \epsfig{figure=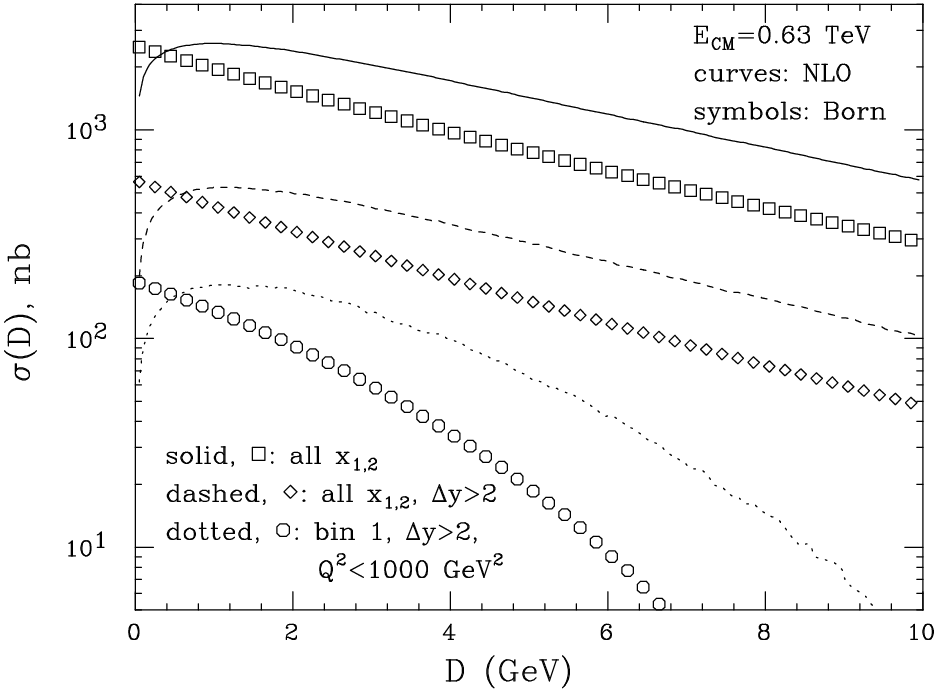,width=0.48\textwidth,clip=}
   \hfill
   \epsfig{figure=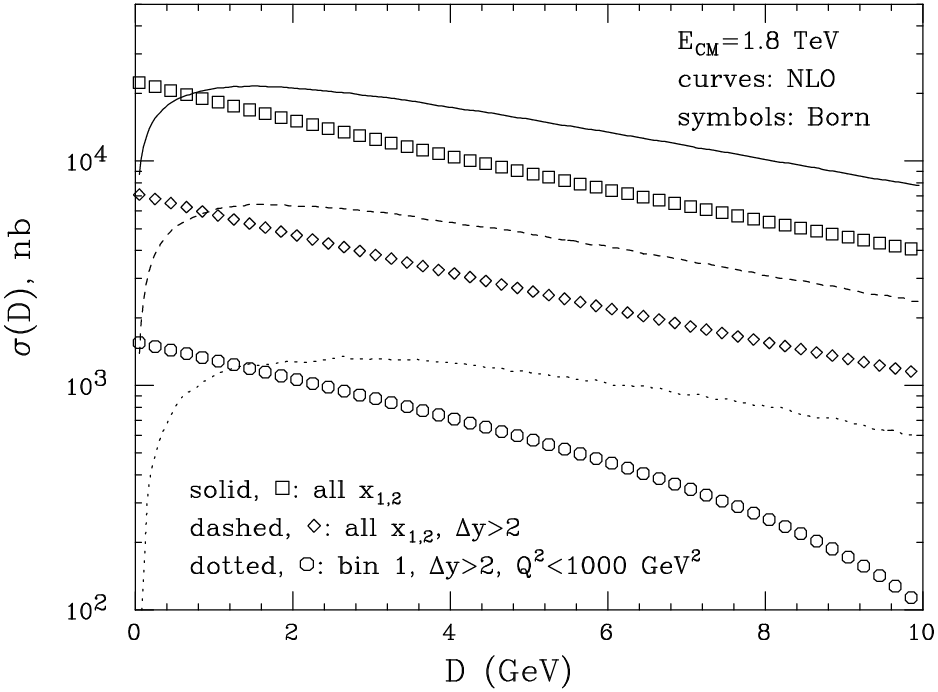,width=0.48\textwidth,clip=} }
\ccaption{}{ \label{fig:xsecs}
Dijet rates, as defined in Eq.~(\protect\ref{ratedef}), for various
cuts $\Cuts$. The cases of $\sqrt{S}=630$~GeV (left) and of
$\sqrt{S}=1800$~GeV (right) are both considered. Dotted curves 
and circles have been rescaled by factor of 10 (left) and 50 (right). 
See the text for details.
}
\end{figure}

The rates defined in Eq.~(\ref{ratedef}) are shown in Fig.~\ref{fig:xsecs},
the left (right) panel presenting the case of $\sqrt{S}=630$~GeV
($\sqrt{S}=1800$~GeV). Each plot consists of three sets of results,
corresponding to different choices of $\Cuts$; for each of these
choices, both the NLO results (displayed by the solid, dashed, and
dotted curves) and the LO results (displayed by the boxes, diamonds
and circles) are given. The solid curves and the boxes are obtained by 
 imposing only the pseudorapidity cuts $\abs{y_i}<3$. The dashed
curves and the diamonds correspond to the previous cuts on $y_i$ plus 
the cut $\dy>2$. Finally, the dotted curves and the circles
are relevant to the cuts given in Eq.~(\ref{Cdef}), plus those 
that define bin number 1 (see Table~\ref{tab:bins}). Notice that
the results relevant to bin 1 have been multiplied by a factor of 10 
and 50 at $\sqrt{S}=630$~GeV and $\sqrt{S}=1800$~GeV respectively, 
so that they can  be shown together with the other results on the same plot.

{}From the cross section definition in Eq.~(\ref{ratedef}), 
it is clear that the smaller $\de$, the
larger the phase space available; thus, one naively expects that the
smaller $\de$, the larger the cross section. This is indeed what
happens at the LO level, regardless of the cuts $\Cuts$.
On the other hand, the NLO cross section increases when $\de$
decreases {\it only if $\de$ is not too close to zero}; when $\de$
 approaches zero, $\sigma(\de)$ has a local maximum, and then
turns over, eventually dropping  below the LO result. As discussed
in Ref.~\cite{Frixione:1997ks}, 
at $\de=0$ the NLO result is finite (i.e., does 
not diverge), but the slope $d\sigma/d\de$ is infinite.
Fig.~\ref{fig:xsecs} thus clearly shows that at $\de=0$ (which
corresponds to the definition adopted in the experimental analysis)
the cross section is affected by large logarithms, that can spoil
the analysis performed in terms of BFKL dynamics, exactly as in
the case of the azimuthal decorrelation.

\begin{table}
\begin{center}
\begin{tabular}{|l||c|c|c||c|c|c|} \hline
& \multicolumn{3}{c||}{NLO} 
& \multicolumn{3}{c|}{Born} 
\\ \hline
& $\de=0$ & $\de=2$ & $\de=4$ 
& $\de=0$ & $\de=2$ & $\de=4$ 
\\ \hline\hline
bin 1
  & -0.12(6) & 1.55(4)  & 2.62(7)  
  & 1.681(3) & 2.340(7) & 4.16(3)  
\\ \hline
bin 2
  & -0.16(4) & 1.14(3)  & 1.46(4)  
  & 1.265(3) & 1.417(4) & 1.739(8) 
\\ \hline
bin 3
  & -0.16(5) & 0.92(3)  & 1.13(4)  
  & 1.074(3) & 1.098(5) & 1.138(6) 
\\ \hline
bin 4
  & -0.19(6) & 0.92(4)  & 1.15(5)  
  & 1.036(4) & 1.045(6) & 1.068(8) 
\\ \hline
bin 5
  & -0.35(5) & 0.82(3)  & 1.01(4)  
  & 1.026(4) & 1.027(6) & 1.020(7) 
\\ \hline
bin 6
  & -0.45(9) & 0.82(7)  & 1.08(9)  
  & 1.015(9) & 1.01(1)  & 1.00(1)  
\\ \hline
\end{tabular} 
\end{center}                                                            
\ccaption{}{\label{tab:ratio}
Fixed-order predictions for the ratio defined in Eq.~(\ref{ratio}).
Numbers in parentheses give the statistical error, which affect
the last digit of the results shown. The values of $\de$ are
given in GeV.
}
\end{table}                                                               
Fig.~\ref{fig:xsecs} already suggests a possible solution to this 
problem: simply define a dijet rate by considering {\it  different}
transverse momentum cuts on the two jets (that is, $\de>0$). From 
the plots, we can expect that the potentially dangerous
logarithms affecting the region $\de=0$ are not large starting
from $\de$ of the order of 3 or 4~GeV. The figure might also
at first sight seem to imply
that  a similar problem  arises in the large $\de$ region in
the case in which the (physically relevant) cuts on $x_1$ and
$x_2$ are imposed (dotted curves and circles). However, it
is easy to understand that in such a case the large difference between
the NLO and LO results is simply due to phase space: in fact, at
 LO $\de\ge 0$ effectively forces {\em both} jets to have
\mbox{$k_\perp\!>\Ecut+\de$}; at NLO, this is no longer true.

Let us therefore consider again the ratio of Eq.~(\ref{ratio}), now rewritten to
 indicate explicitly the cuts adopted:
\beq
R(\de,\Cuts)=\sigma(\de,\Cuts;\sqrt{S}=1800~{\rm GeV})/
             \sigma(\de,\Cuts;\sqrt{S}=630~{\rm GeV})\, ,
\label{ratioFO}
\eeq
with $\Cuts$ given in Eq.~(\ref{Cdef}), and additional
 (binning) cuts on $x_1$ and $x_2$.
Our predictions for $R$, both at NLO and  LO, are presented
in Table~\ref{tab:ratio}, where we show the results for all of the bins
of Table~\ref{tab:bins}. The entries relevant to $\de=0$ display
a pathological (negative)  behaviour at NLO. However, even if
they were positive, they could not be considered reliable, since
any fixed-order QCD computation (beyond LO) is unable to give a sound 
prediction in this case. On the other hand, we see that for larger
values of $\de$ the situation improves, in the sense that it reproduces
our naive expectation: the ratio should converge towards one, for
increasing $\dy$ (i.e., larger bin numbers); while
for $\de=2$~GeV the NLO results are still sizeably different from the
LO results, in the case of $\de=4$~GeV the NLO and LO results are 
statistically compatible (within one standard deviation) for bins 3--6, 
and they are both approaching one.

Inspection of Fig.~\ref{fig:xsecs} and Table~\ref{tab:ratio} tells us
that, in order to avoid the presence of large logarithms of non-BFKL
nature in the cross section, a value of $\de=4$~GeV is probably
a better choice than $\de=2$~GeV. Of course, the larger $\de$,
the smaller the cross section, and therefore the fewer the events.
In order to give an estimate of the loss of events that one faces
when going from $\de=0$ to larger values, we present in 
Table~\ref{tab:xsec} our LO predictions for the rate defined
in Eq.~(\ref{ratedef}), with the cuts of Eq.~(\ref{Cdef}) and
our six bins. Of course, it is well known that NLO corrections are
mandatory in jet physics to get good agreement with data. However,
here we just want to have a rough idea of the number of events lost
when increasing one of the transverse energy cuts; this number
is sensibly predicted by the ratio \mbox{$\sigma(\de)/\sigma(\de=0)$},
even if $\sigma$ is only computed at LO. From 
the table, we see that at $\de=4$~GeV the number of events
decreases, compared to the case $\de=0$, by a factor
slightly larger than two; this factor gets much larger only for
the first two bins at $\sqrt{S}=630$~GeV, which are however less
relevant from the point of view of BFKL dynamics.
\begin{table}
\begin{center}
\begin{tabular}{|l||c|c|c||c|c|c|} \hline
& \multicolumn{3}{c||}{$\sqrt{S}=1.8$~TeV} 
& \multicolumn{3}{c|}{$\sqrt{S}=0.63$~TeV} 
\\ \hline
& $\de=0$ & $\de=2$ & $\de=4$ 
& $\de=0$ & $\de=2$ & $\de=4$ 
\\ \hline\hline
bin 1
  & 31.01(3) & 21.24(3) & 14.19(2) 
  & 18.46(3) & 9.080(2) & 3.399(1) 
\\ \hline
bin 2
  & 22.66(2) & 15.52(2) & 10.36(2) 
  & 17.91(2) & 10.98(2) & 5.969(2) 
\\ \hline
bin 3
  & 13.50(2) & 9.22(2)  & 6.16(2)  
  & 12.57(2) & 8.39(2)  & 5.41(2)  
\\ \hline
bin 4
  & 12.11(3) & 8.29(2)  & 5.54(2)  
  & 11.69(3) & 7.20(2)  & 5.19(2)  
\\ \hline
bin 5
  & 7.19(1)  & 4.90(1)  & 3.26(1)  
  & 7.10(1)  & 4.78(2)  & 3.19(1)  
\\ \hline
bin 6
  & 4.25(2)  & 2.89(2)  & 1.92(1)  
  & 4.19(2)  & 2.85(2)  & 1.92(1)  
\\ \hline
\end{tabular} 
\end{center}                                                            
\ccaption{}{\label{tab:xsec}
Cross sections in nanobarns as given in Eq.~(\ref{ratedef}), at the LO and for
two different centre-of-mass energies. Statistical errors are
given in parentheses.
}
\end{table}                                                               

\subsection{Dijet production in the BFKL theory with an asymmetric cut}
\label{sec:BFKLcuts}

\begin{figure}[htb]
\centerline{
        \epsfig{file=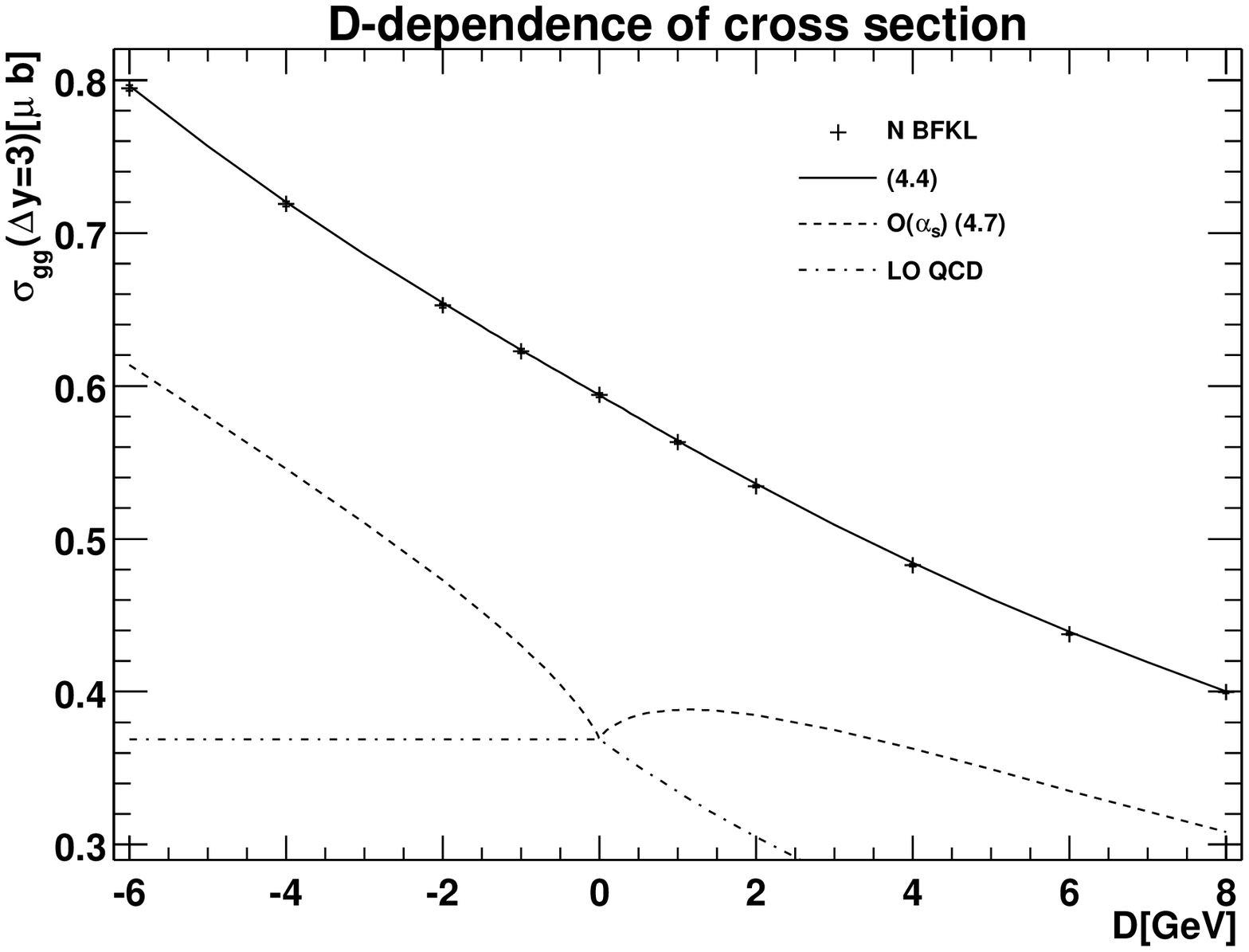,width=13cm} }
\ccaption{}{ \label{ggxsecasy}
The dependence of the gluon-gluon subprocess cross section 
on the offset $\de$, for fixed separation $\Delta y = 3$. 
The resummed prediction Eq.~(\ref{qmaxdelta}) is shown as a solid line, with the
results of the corresponding BFKL Monte Carlo calculation superimposed.
The dash-dotted line is the LO contribution, and the 
dashed line is the ${\cal O}(\alpha_s)$ contribution of 
Eq.~(\ref{expdeltasol}). The parameter 
values are $\alpha_s = 0.164$, $ \Ecut = 20$~GeV, 
$Q^2_{\rm max} = \infty$.
}
\end{figure}

We now turn again to the BFKL equation, and study the dependence
on the offset $\de$ introduced in the previous subsection, in 
essentially the same way as we did in 
Section~\ref{sec:naiveBFKL}. We start by 
integrating the gluon-gluon cross section (\ref{cross}) over
$\kta\!>\!\Ecut$ and $\ktb\!>\!\Ecut + \de$ with the
upper cut \mbox{$Q^2=\kta\ktb\!\!<1000$~GeV$^2$}, and with the $x$'s
defined as in the Mueller-Navelet analysis, \eqn{mnkin},
\beqa
\lefteqn{
\hat\sigma_{gg}(\kta\!>\!\Ecut,\ktb\!>\!\Ecut+\de,
\kta\ktb\!\!< Q^2_{\rm max}) 
= {\pi C_A^2\alpha_s^2\over 2\Ecut (\Ecut+\de)} } 
\label{qmaxdelta}
\\*
&\times& \Biggl\{ \cF\left(\dy,{\Ecut\over \Ecut+\de}\right)
- {\Ecut (\Ecut+\de)\over 2 Q^2_{\rm max}}
\left[ \cG\left(\dy,{\Ecutt\over Q^2_{\rm max}}\right) +
\cG\left(\dy,{(\Ecut+\de)^2\over Q^2_{\rm max}}\right)\right] 
\Biggr\}\, ,\nonumber
\eeqa
with $\cF$ and $\cG$ defined in~\eqn{ffunct} and~\eqn{gfunct}.
Repeating the calculation as in \sec{sec:twotwo}, 
with the $x$'s defined as in \eqn{nkin0},
yields a cross section of the same form as \eqn{qmaxdelta} up
to replacing the rapidity interval (\ref{mnrap}) with the constant
(\ref{const}), the upper bound $Q^2_{\rm max}$ above with \eqn{qmax2} and 
the function $\cF$ ($\cG$) with $\cFt$ ($\cGt$), \eqn{ftfunct} 
(\eqn{gtfunct}). At $\de = 0$ we recover \eqns{qmaxnaive}{qmaxcomb} 
respectively. However, near $\de = 0$ \eqn{qmaxdelta} and its analogous one 
with the tilda functions display the same qualitative behaviour when
expanded to NLO as the exact NLO cross section~\cite{Frixione:1997ks}. 
In order to see this, we take \eqn{qmaxdelta} in the limit 
$Q^2_{\rm max}\to \infty$, such that only the first term 
on the right hand side of \eqn{qmaxdelta} survives. We analyse its NLO term 
by expanding its exponential to ${\cal O}(\alpha_s)$,
\beqa
\lefteqn{\hat\sigma_{gg}(\kta\!>\!\Ecut,\ktb\!>\! 
\Ecut + \de)} 
\label{expsol}
\\*\nonumber &=&
{C_A^2\alpha_s^2\over 4 \Ecut (\Ecut+ \de)}
\int_{-\infty}^{\infty} d\nu\, 
{ 1 - 2{\bar \alpha_s} \dy\, \left[
{\rm Re}\,\psi \left({1/2} +i\nu \right) - \psi(1) \right]
\over \nu^2 + 1/4}\, \left(\Ecut\over \Ecut+ \de\right)^{2i\nu}
+{\cal O}\left( (\bar\alpha_s\dy)^2 \right)\,.
\eeqa
The denominator has poles at $\nu = \pm i/2$. For the LO term, the integration
over $\nu$ is straightforward. For the NLO term, we use the integral
representation of the digamma function,
\beq
\psi(z) - \psi(1) = \int_0^1 dx\ {1-x^{z-1} \over 1-x}
\label{psif}
\eeq
and after performing the integrals over $\nu$ and $x$, we find
\beqa
\lefteqn{\hat\sigma_{gg}(\kta\!>\!\Ecut,\ktb\!>\! 
\Ecut + \de)} 
\label{expdeltasol}
\\*\nonumber &&\phantom{\,}
={\pi C_A^2\alpha_s^2\over 2 } \Biggl\{ {1\over \Ecutt}
\left[ 1 - {\bar \alpha_s} \dy\, \left( {-2\Ecut \de - \de^2
\over (\Ecut + \de)^2} \ln{ -2\Ecut \de - \de^2 \over \Ecutt}
+ 2 \ln{\Ecut+ \de\over \Ecut} \right) \right] \theta(-\de) 
\\*\nonumber &&\phantom{\,}
+{1\over (\Ecut + \de)^2}
\left[ 1 - {\bar \alpha_s} \dy\, \left( {2\Ecut \de + \de^2
\over \Ecutt} \ln{ 2\Ecut \de + \de^2 \over (\Ecut+ \de)^2}
+ 2 \ln{\Ecut\over \Ecut+ \de} \right) \right] \theta(\de) \Biggr\} 
\\*\nonumber &&\phantom{\,}
+{\cal O}\left( (\bar\alpha_s\dy)^2 \right)\,.
\eeqa
At LO, transverse-momentum conservation forces the cross section to 
behave like \mbox{$1/\Ecutt$ for $\de < 0$} and like 
$1/(\Ecut+ \de)^2$ for $\de > 0$, even though the cuts over the transverse
momenta are asymmetric, see Fig.~\ref{ggxsecasy}. 
For $\de=0$, \eqn{expdeltasol} reduces to
\eqn{ysmallsol}. For $|\de|\ll \Ecut$, \eqn{expdeltasol} becomes
\beq
\hat\sigma_{gg}(\kta\!>\!\Ecut,\ktb\!>\!\Ecut + \de)\ =\
{\pi C_A^2\alpha_s^2\over 2 \Ecutt}
\left[ 1 - {\bar \alpha_s} \dy\, {2|\de| 
\over \Ecut} \ln{ 2|\de| \over \Ecut}
+ {\cal O}(\de)  \right]
+{\cal O}\left( (\bar\alpha_s\dy)^2 \right)\,.
\label{smalldeltasol}
\eeq
The slope of \eqn{expdeltasol} with respect to $\de$ is negative (positive)
for $\de < 0\, (> 0)$,
and infinite at $\de = 0$, in agreement with Ref.~\cite{Frixione:1997ks}.
In addition, by using \eqn{smalldeltasol} to evaluate the 
ratio~(\ref{ratio}) and
remembering that asymptotically $\dy_{_A} \to \dy_{_B} + 
2\ln(1800/630)$, we find that  the NLO BFKL ratio also goes to 1 as $\dy$ 
grows, in agreement with \tab{tab:ratio}. 

In the BFKL Monte Carlo approach, the implementation of asymmetric
cuts on the jets is straightforward. For fixed $\alpha_s$ and no additional
cuts or parton densities, the analytic result of Eq.~(\ref{qmaxdelta}) 
is reproduced, see \fig{ggxsecasy}.

Finally, we can use the BFKL Monte Carlo to calculate the `D0' cross section
ratios defined in Eq.~(\ref{ratio}), in the various bins. 
Table~\ref{tab:BFKLratios} gives the predictions
using the Monte Carlo run in two modes:
\begin{description}
\item[{\bf Naive:}] fixed $\alpha_s$, no kinematic constraints, parton
  densities evaluated at Bjorken $x$'s given in Eq.~(\ref{nkin0}).
\item[{\bf Full:}] running $\alpha_s$, energy--momentum conservation applied, 
  parton densities evaluated at $x_a$, $x_b$ values given in 
  Eq.~(\ref{emcons}).
\end{description}
Evidently neither the naive BFKL nor the BFKL MC calculation shows 
the `pathological' behaviour of the exact NLO calculation 
 at $\de =0$ (this is already
apparent from \fig{ggxsecasy}). Instead the numbers are quite stable against
variations in $\de$. For all $\de$ the naive BFKL calculation shows 
an initial decrease in
the cross section ratios, before reaching a minimum around bin 4 where the
expected rise due to BFKL dynamics sets in. The initial decrease is simply
the subasymptotic effect of the $\Delta y>2$ cut on the cross section at 
$\sqrt{s}=630$ GeV, as discussed in Section~\ref{sec:twotwo}, 
and consistent with the qualitative
behaviour of the Born cross section ratios in Table~\ref{tab:ratio}.
On the other hand, if the effects from the parton densities did 
factorize out completely, one would expect an asymptotic (in bin number)
ratio of $R=(S_A/S_B)^\lambda$ with $\lambda=\frac{\alpha_sC_A}{\pi}4\ln
2\approx 0.45$~~\cite{Orr:1998hc}. This gives $R\approx 2.6$ for the D0 
values. 
However, we have already argued that such a  rise is {\it not}
expected in the D0 analysis, mainly because of the rather stringent
$Q^2$ cut (see \fig{ggxsec} and \fig{d0ggxsec}). 
Furthermore, the expected rise is also slightly
decreased by the cut in $x_a,x_b\le 1$, and hence on $\kta$ and $\ktb$, 
introduced by the parton densities.
The full BFKL MC calculation ratios also show an
initial decrease to a minimum around bin 4. However now the ratio is {\it below} 
1 already from bin 3 onwards. Such an effect was already reported
and explained 
in Ref.~\cite{Orr:1998hc}. It is a kinematic effect due to  an effective upper limit on the 
transverse momentum allowed for each emitted gluon. As the rapidity separation between
the dijets is increased towards its maximum allowed value, the BFKL gluon phase space
is squeezed from above and the `naive' cross section is heavily suppressed. The higher
the collision energy the more dramatic the effect, and hence the ratio $R$ 
falls below 1.  

\begin{table}
\begin{center}
\begin{tabular}{|l||c|c|c||c|c|c|} \hline
& \multicolumn{3}{c||}{Naive BFKL} 
& \multicolumn{3}{c|}{BFKL MC} 
\\ \hline
& $\de=0$ & $\de=2$ & $\de=4$ 
& $\de=0$ & $\de=2$ & $\de=4$ 
\\ \hline\hline
bin 1
& 2.16(4) &2.56(5) & 3.61(8)
& 1.615(3)& 2.013(5)& 2.922(8)
\\ \hline
bin 2
& 1.47(4) & 1.53(4)& 1.73(4)
& 1.048(2)& 1.114(2)& 1.289(3)
\\ \hline
bin 3
& 1.22(4) &1.16(4) & 1.10(3)
& 0.866(2)& 0.851(2)& 0.872(2)
\\ \hline
bin 4
& 1.18(4) &1.12(4) & 1.14(4)
& 0.806(4)& 0.783(2)& 0.787(2)
\\ \hline
bin 5
& 1.26(6) & 1.14(5)& 1.11(5)
& 0.847(2)& 0.824(2)& 0.820(3)
\\ \hline
bin 6
& 1.31(8) & 1.28(7)& 1.20(6)
& 0.863(3)& 0.841(3)& 0.838(3)
\\ \hline
\end{tabular} 
\end{center}                                                            
\ccaption{}{\label{tab:BFKLratios}
BFKL predictions for the ratio defined in Eq.~(\ref{ratio}).
Numbers in parentheses give the statistical error, which affect
the last digit of the results shown. The values of $\de$ are
given in GeV.
}
\end{table}                                                               

\section{Conclusions}
\label{sec:concl}

In this paper, we have reconsidered the suggestion by Mueller and Navelet
of studying dijet cross sections at large rapidity intervals and for
different hadronic centre-of-mass energies, in order to find evidence
of BFKL physics. We were motivated by a recent paper
by the D0 Collaboration~\cite{Abbott:2000ai}, where dijet data have
been used to measure the effective BFKL intercept by comparison
with  the standard analytic asymptotic formulae given by Mueller and 
Navelet. In fact, we have argued that the definition of the momentum 
fractions used by D0 and 
some of the acceptance cuts imposed by D0 spoil the correctness
of this procedure, and require a more careful theoretical investigation.

In particular, we are concerned by a difference between D0 and the standard
Mueller-Navelet analysis in the reconstruction of the momentum fraction $x$ of 
the incoming partons, by the presence of an upper bound on
the momentum transfer $Q^2$, and by the requirement that the two tagged
jets have the same minimum transverse energy.

The $Q^2$ cut allows, at the experimental level, and together with 
the binning cuts on $x_{1,2}$, a reduction of the systematic errors,
since in the ratio of the cross sections measured at 
different centre-of-mass energies 
the dependence on the parton densities cancels to a 
significant extent. We have shown that, at the level of partonic cross 
sections, the upper bound on $Q^2$ and the $x$'s used in the D0 analysis
reduce the Mueller-Navelet cross section
by a factor of more than 5. On the other hand, the dependence on such 
a cut, as well as the dependence on the precise definition of the $x$'s,
cancel out when considering the ratio of cross sections obtained
at different energies. However, this is only true when the {\em asymptotic}
forms of the cross sections are considered. Unfortunately, at the energies 
and rapidity intervals probed at the Tevatron, it appears that the 
asymptotic expansions do not reproduce accurately enough the exact 
analytic results; in particular, the quality of the approximations 
are worse in the case in which an upper cut on $Q^2$ is imposed.
We are therefore led to conclude that, regardless of the use of
cross sections or of rates of cross sections to study BFKL physics,
the effect of an upper bound on $Q^2$ cannot be ignored.

As far as the cuts on the transverse momenta of the trigger jets
are concerned, we have pointed out that in the case in which such
cuts are chosen to be equal, the cross sections are plagued with 
large logarithms of perturbative, non-BFKL origin. In this sense, the total
dijet rates are therefore on the same footing as the azimuthal
decorrelations. We therefore believe that a much safer choice
is to have different cuts on the transverse momenta of
the two jets.

\section*{Acknowledgements}
V.D.D. and S.F. would like to thank CERN TH Division for the hospitality
while this work was performed. JRA acknowledges the financial support of the
The Danish Research Agency. The work of C.R.S. was supported by the 
US National Science Foundation under grants PHY-9722144 and PHY-0070443.

\end{document}